\documentclass[12pt]{article}

\title{Theoretical estimates of the logarithmic phonon spectral moment 
       for monatomic liquids}

\author{Eric D.\ Chisolm and Duane C.\ Wallace \\ Theoretical Division, 
        Los Alamos National Laboratory \\ Los Alamos, NM~~87545}

\usepackage{graphicx}

\begin{document}

\maketitle

\vspace*{-3.0in} \begin{flushright} LA-UR-03-5331 \end{flushright}
\vspace*{2.5in}

\begin{abstract}
We calculate the logarithmic moment of the phonon frequency spectrum
at a single density for 29 monatomic liquids using two methods, both
suggested by Wallace's theory of liquid dynamics: The first method
relies on liquid entropy data, the second on neutron scattering data
in the crystal phase.  This theory predicts that for a class of
elements called ``normal melters,'' including all 29 of these
materials, the two estimates should closely match, and we find that
they agree to within a few percent.  We also perform the same
calculations for four ``anomalous melters,'' for which we expect the
two estimates to differ markedly; we find that they disagree by
factors almost up to three.  From our results we conclude that the
liquid entropy estimates of the logarithmic moment, applicable both to
normal and anomalous melters, are trustworthy to a few percent, which
makes them reliable for use in estimates of various liquid transport
coefficients.
\end{abstract}

\section{Introduction}
\label{intro}

Moments of the phonon frequency distribution of systems in the liquid
phase appear to set the timescales on which certain correlation
functions decay, thus affecting the values of the transport
coefficients to which they correspond by Green-Kubo relations
\cite{diff,march}; so determining these coefficients requires the
reliable estimation of the relevant moments.  The comprehensive theory
of monatomic liquids proposed by Wallace (see \cite{liq} for a review)
allows one to estimate one of these moments, the logarithmic moment
$\Theta_0^l$, from experimental measurements of the liquid entropy,
but for these applications one would like greater confidence that this
estimate is a reliable one.  In the first paper on this theory
\cite{wall1}, it was shown that the resulting expression for entropy
together with the estimate $\Theta_0^l \approx \Theta_0^{\rm c}$,
where $ \Theta_0^{\rm c}$ is the corresponding moment in the crystal
phase (this estimate is explained below) reproduced the temperature
dependence of the entropy for six elements to an accuracy of roughly
$2\%$ from melt to three times the melting temperature, supporting
both the general form for the entropy and the approximation for
$\Theta_0^l$.  However, for most materials entropy data well above
melting are unavailable, and one must ask a different question: If one
computes $\Theta_0^l$ from only a single entropy data point near
melting, how reliable is the result?  We will argue that this estimate
is very good by considering data for 29 ``normal melters'' (elements
like the six in \cite{wall1}, for which the theory claims
$\Theta_0^{\rm c} \approx \Theta_0^l$ is a good approximation),
showing that $\Theta_0^l$ computed from entropy data at the density of
the liquid at melt at 1 bar closely matches $\Theta_0^{\rm c}$
computed at the same density from neutron scattering data and other
sources.  The internal consistency of these two ways of estimating
$\Theta_0^l$ for normal melters, both provided by the same theory but
utilizing independent sources of data, increases our confidence that
the entropy estimates of $\Theta_0^l$ will be trustworthy even for
elements that are not normal melters, and thus aren't expected to
satisfy the $\Theta_0^{\rm c} \approx \Theta_0^l$ approximation.

In Section \ref{theory} we use this theory to deduce the two ways of
estimating $\Theta_0^l$, one of which applies generally and one of
which applies only to normal melters, and in Section \ref{exp} we use
experimental data to compute the two estimates, finding that they
match to $7\%$.  Finally, in Section \ref{concl} we consider the
significance of these results.

\section{Two estimates of $\Theta_0^l$ for normal melters}
\label{theory}

The theory of liquid dynamics that we use here was developed in
response to two trends in the experimental data for monatomic liquids
at melt \cite{liq}: (1) The nuclear contribution to the specific heat
tends to lie near $3k$ per atom in both crystal and liquid phases; and
(2) the entropy of melting data at constant density naturally divide
the liquids into two classes: the ``normal melters,'' for which the
entropy clusters around $0.8k$ per atom, and the ``anomalous
melters,'' for which the entropy of melting is much higher and which
undergo significant electronic structure change (e.g., semimetal
crystal to metal liquid) upon melting.  The specific heat results
suggest that the liquid undergoes collective harmonic motion much
as the crystal does, and the entropy of melting data are naturally
interpreted to show that the liquid, as opposed to the crystal, moves
among $w^N$ potential valleys, where $N$ is the number of atoms and
$\ln w \approx 0.8$, thus contributing an extra configurational
entropy on melting, with the excess for the anomalous melters being
due to the electronic structure change.  These observations led to two
hypotheses:
\begin{enumerate}
\item The potential landscape in which the atoms move in the liquid
phase is dominated by approximately $w^N$ intersecting nearly-harmonic
valleys, and the atomic motion is a combination of oscillations in
one of these valleys and nearly instantaneous transitions from one
valley to another called {\em transits}.  Transits are responsible for
diffusion.
\item The valleys fall into three categories: the few {\em
crystalline} valleys in which the system exists as a single crystal,
the more numerous {\em symmetric} valleys corresponding to
configurations that retain some of the crystal symmetry, and the
overwhelmingly numerically dominant {\em random} valleys, which retain
no remnant symmetry and which all have the same depth and shape (and
thus the same phonon frequency spectrum) in the large $N$ limit.
\end{enumerate}
These hypotheses have been extensively tested (as summarized in
\cite{liq}) and can be used to calculate thermodynamic quantities such
as entropy per atom; in the classical limit, where most liquids
reside, the result is (Section 4 of \cite{liq})
\begin{equation}
S^l =  k\ln w + 3k\left[\ln(T/\Theta_0^l) + 1\right] + S_{anh}^l + S_{bdy}^l 
       + S_{el}^l + S_{mag}^l,
\label{Sl}
\end{equation}
where $S_{anh}^l$, $S_{bdy}^l$, $S_{el}^l$, and $S_{mag}^l$ are the
contributions to the entropy from anharmonicity, the presence of
boundaries of the potential valleys, the thermal excitation of the
electrons, and magnetic effects.  (The anharmonic and boundary
contributions are included in a single term, and the magnetic
contribution is not included at all, in Section 4 of \cite{liq}, but
$S_{mag}$ is discussed on pp.\ 202-203 of \cite{book}.)  Thus, if one
can compute $S_{anh}^l$, $S_{bdy}^l$, $S_{el}^l$, and $S_{mag}^l$ at a
given density (or argue that they are negligibly small), one can
calculate $\Theta_0^l$ from liquid entropy data at that density at a
single temperature (or a range of temperatures, if data are
available).

While the theory argues that this means of calculating $\Theta_0^l$ is
valid for all monatomic liquids, it also makes another prediction for
the value of $\Theta_0^l$ which applies only to the normal melters,
namely that $\Theta_0^l$ is nearly equal to $\Theta_0^{\rm c}$ at the
same density.  The prediction comes about as follows (see Subsection
4.4 of \cite{liq}): The entropy per atom of the crystal phase in the
classical limit is calculated from lattice dynamics to be
\begin{equation}
S^{\rm c} = 3k\left[\ln(T/\Theta_0^{\rm c}) + 1\right] + 
            S_{anh}^{\rm c} + S_{el}^{\rm c} + S_{mag}^{\rm c},
\end{equation}
where $S_{anh}^{\rm c}$, $S_{el}^{\rm c}$, and $S_{mag}^{\rm c}$
represent the contributions from anharmonicity, the electrons, and
magnetic effects, respectively, so the entropy of melting at constant
density $\Delta S$ is given by
\begin{eqnarray}
\Delta S & = & S^l(T_m) - S^{\rm c}(T_m) \nonumber \\
         & = & k\ln w + 3k\ln(\Theta_0^{\rm c}/\Theta_0^l) + \left(S_{anh}^l -
               S_{anh}^{\rm c}\right) + \nonumber \\
         &   & \left(S_{el}^l - S_{el}^{\rm c}\right) + 
               \left(S_{mag}^l - S_{mag}^{\rm c}\right).
\end{eqnarray}
$S_{bdy}^l$ is omitted here because boundary effects are
non-negligible only at temperatures well above melting.  For a normal
melter, the lack of significant change in electronic structure
suggests that $S_{el}^l \approx S_{el}^{\rm c}$, and assuming that
anharmonic and magnetic effects are small (or at least comparable in
the two phases), we predict $\Delta S \approx k\ln w +
3k\ln(\Theta_0^{\rm c}/\Theta_0^l)$.  Experiment shows that for the
normal melters $\Delta S = 0.8k$ with a small scatter, which strongly
suggests that $\ln w = 0.8$ (since $w$ is the only available parameter
that is not strongly material-dependent) and that $\Theta_0^{\rm c}
\approx \Theta_0^l$, with anharmonicity and small differences between
the two values of $\Theta_0$ accounting for the scatter.  Thus one
should be able to estimate $\Theta_0^l$ for normal melters reasonably
closely by determining $\Theta_0^{\rm c}$ from appropriate data.

If this interpretation of the entropy of melting data is correct,
these two estimates of $\Theta_0^l$ for normal melters should
approximately match, which will in turn increase our confidence in
estimates of $\Theta_0^l$ from liquid entropy data for any liquid,
normal or anomalous.  Determining whether they do in fact match is the
subject of the next Section.

\section{Comparison of experimental estimates}
\label{exp}

Let $\rho_{lm}$ be the density of the liquid at melt at 1 bar (with
the exception of Ar; see the Appendix).  We will compare
$\Theta_0^{\rm c}(\rho_{lm})$ and $\Theta_0^l(\rho_{lm})$ for 29
normal melters for which accurate data are available.  

To compute $\Theta_0^{\rm c}(\rho_{lm})$, we consult neutron
scattering or other data which give us $\Theta_0^{\rm c}(\rho_{
meas})$, where $\rho_{meas}$ is some density not too far from
$\rho_{lm}$, and $\gamma^*$, the high-temperature Gr\"{u}neisen
parameter.  From these two quantities one can calculate $\Theta_0^{\rm
c}(\rho_{lm})$ from the relation
\begin{equation}
\Theta_0^{\rm c}(\rho_{lm}) = \Theta_0^{\rm c}(\rho_{meas}) \left(
                              \frac{\rho_{lm}}{\rho_{meas}}\right)^{\gamma^*}, 
\label{theta0c}
\end{equation}
which is highly accurate as long as $|\rho_{lm} - \rho_{meas}|$ is not
too large.

To compute $\Theta_0^l(\rho_{lm})$ from liquid entropy data, we first
define the harmonic contribution to the entropy $S_{harm}^l$ by
\begin{equation}
S_{harm}^l = k\ln w + 3k\left[\ln(T/\Theta_0^l) + 1 + 
             \frac{1}{40}\left(\frac{\Theta_2^l}{T}\right)^2\right],
\end{equation}
where $\Theta_2^l$ is the quadratic moment of the liquid phonon
distribution.  This is the same as the first two terms in Eq.\
(\ref{Sl}) for the classical limit of the liquid entropy, carried to
the next higher order term in $T^{-2}$.  Then if $S_{expt}^l$ is the
experimental value of the liquid entropy at the melting temperature
$T_m$ at 1 bar,
\begin{equation}
S_{expt}^l =  S_{harm}^l(\rho_{lm}, T_m) + S_{anh}^l + S_{el}^l + S_{mag}^l
\label{Sliq}
\end{equation}
(again, $S_{bdy}^l$ is negligible at melt), so to determine $S_{harm}^l$
(and thus $\Theta_0^l$) we need $\Theta_2^l$ and the remaining three
terms on the right hand side of Eq.\ (\ref{Sliq}).  The standard
approximation $\Theta_2^l = e^{1/3} \Theta_0^l$ is perfectly
reasonable here, since the $\Theta_2^l$ term contributes less than
$1\%$ to the entropy at the values of $T_m$ we consider, and we argue
that the anharmonic and magnetic terms in the liquid at melt will be
roughly equal to their counterparts in the crystal just before melt.
(Recall that these two terms can be calculated for the crystal,
because we have independent data for $\Theta_0^{\rm c}$ and thus we
can determine the harmonic contribution to the crystal entropy
directly; once the electronic term is also found from theory, the
anharmonic and magnetic parts are simply what's left.)  The electronic
term can be calculated different ways for different materials; see the
Appendix for a discussion of how $S_{el}^l$ was calculated for the
materials here.

\begin{figure}
  \centering
  \includegraphics[height=0.5\textheight]{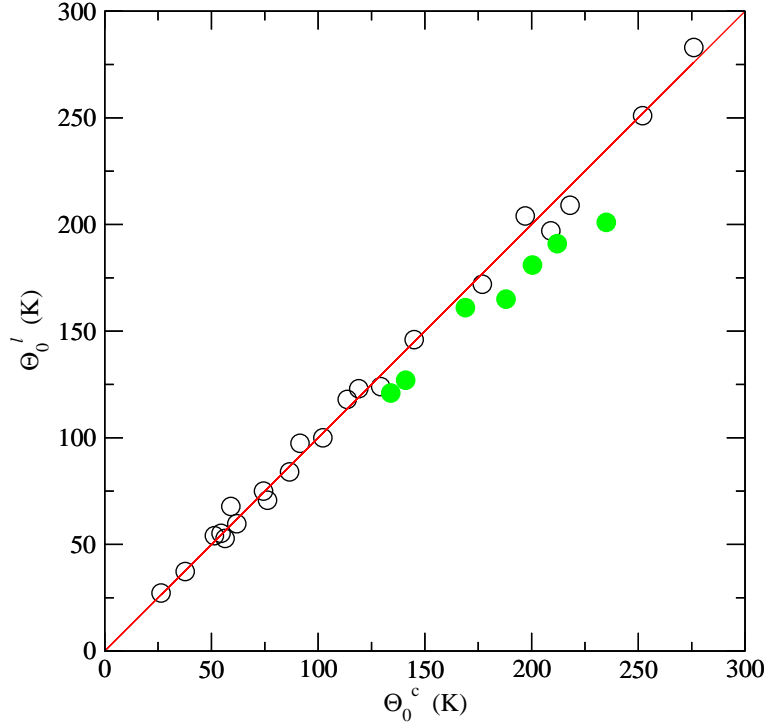}
  \caption{$\Theta_0^l(\rho_{lm})$ versus $\Theta_0^{\rm c}(\rho_{lm})$ 
           for the 29 elements in Table \ref{tabthetas}.  The seven elements 
           indicated by solid circles, all but one of which are transition 
           metals, deviate systematically from the line $\Theta_0^l(\rho_{lm})
           = \Theta_0^{\rm c}(\rho_{lm})$; we believe this is due either to
           these materials being weakly anomalous melters or to uncertainties
           in the entropy data.}
  \label{figthetas}
\end{figure}

These calculations were performed for 29 normal melters, using the
data collected in the Appendix, and the results are shown in Table
\ref{tabthetas}.
\begin{table}
\centering
\begin{tabular}{lccc} \hline\hline
Element & $\Theta_0^{\rm c}$ (K) & $\Theta_0^l$ (K) & 
$\Theta_0^l / \Theta_0^{\rm c}$ \\ \hline
Cs & 26.4   & 27.2  & 1.03   \\
Cd & 86.6   & 84    & 0.97   \\
Tl & (54.5) & 55.2  & (1.01) \\
Li & 252    & 251   & 1.00   \\
Na & 102.2  & 100   & 0.98   \\
K  & 61.9   & 59.7  & 0.96   \\
Rb & 37.7   & 37.3  & 0.99   \\
Mg & 200.4  & 181   & 0.91   \\
Sr & (76.3) & 70.7  & (0.93) \\
Cu & 176.9  & 172   & 0.97   \\
Ag & 113.6  & 118   & 1.04   \\
Au & 91.5   & 97.4  & 1.06   \\
Zn & 129.3  & 124   & 0.96   \\
Hg & 56.4   & 52.8  & 0.94   \\
Al & 209.0  & 197   & 0.94   \\
In & 74.4   & 75.0  & 1.01   \\
Pb & 51.4   & 54.1  & 1.05   \\
Ar & 59.1   & 67.8  & 1.15   \\ 
V  & 212    & 191   & 0.90   \\
Nb & 169    & 161   & 0.95   \\
Ta & 134    & 121   & 0.90   \\
Pt & 119    & 123   & 1.03   \\
Pd & 145    & (146) & (1.01) \\
Ni & 218    & 209   & 0.96   \\
Cr & (276)  & (283) & (1.03) \\
Mo & 235    & (201) & (0.86) \\
W  & 188    & (165) & (0.88) \\
Ti & (197)  & 204   & (1.04) \\
Zr & (141)  & 127   & (0.90) \\ \hline \hline
\end{tabular}
\caption{$\Theta_0^{\rm c}(\rho_{lm})$ and $\Theta_0^l(\rho_{lm})$ for
29 normal melters, together with their ratios.  Less reliable numbers
are shown in parentheses; see the Appendix for a discussion of the
reliability of data.}
\label{tabthetas}
\end{table}
In Figure \ref{figthetas}, $\Theta_0^l(\rho_{lm})$ is plotted against
$\Theta_0^{\rm c}(\rho_{lm})$ for all 29 elements, along with the line
$\Theta_0^l(\rho_{lm}) = \Theta_0^{\rm c}(\rho_{lm})$ for comparison.
First, we note that all 29 points cluster around the line to good
approximation; second, we note that most of the scatter is due to the
systematic deviation of the seven materials indicated by solid circles,
all but one of which (Mg) are transition metals.  This could be due
either to the larger uncertainty in the experimental entropy results
for the transition metals or to the fact that some of these materials
are weakly anomalous melters (the entropy of melting data for Mo and W
particularly suggest this \cite{book}).
Our data for all 29 materials show
\begin{equation}
\frac{\Theta_0^l}{\Theta_0^{\rm c}} = 0.978 \pm 0.063
\end{equation}
at $\rho_{lm}$, or equivalently that the rms deviation of $\Theta_0^l
/ \Theta_0^{\rm c}$ from $1$ is $0.067$ at $\rho_{lm}$.  As we explain
in the Appendix, given the quality of our data we expect roughly 3\%
(at most 4\%) errors in this ratio; so as the theory predicts, the two
are equal to a high degree of approximation, and about half of the
difference is accounted for by experimental error.

In contrast, if we consider analogous calculations for four anomalous
melters, we find the results shown in Table \ref{tabanom}.
\begin{table}
\centering
\begin{tabular}{lccc} \hline\hline
Element & $\Theta_0^{\rm c}$ (K) & $\Theta_0^l$ (K) & 
$\Theta_0^l / \Theta_0^{\rm c}$ \\ \hline
Ga & 169  & 100 & 0.59 \\
Si & 436  & 162 & 0.37 \\
Ge & 255  & 91  & 0.36 \\
Sn & 94.3 & 74  & 0.78 \\ \hline\hline
\end{tabular}
\caption{$\Theta_0^{\rm c}(\rho_{lm})$ and $\Theta_0^l(\rho_{lm})$ for
four anomalous melters, together with their ratios.  As expected, none of
the ratios are near unity, reflecting the changes in the interatomic
forces (and thus phonon frequencies) that occur at melt for these
materials.}
\label{tabanom}
\end{table}
The change in electronic structure at melt results in great changes in
the interatomic forces, resulting in large changes in the phonon
frequencies.  For these materials, the approximation $\Theta_0^l =
\Theta_0^{\rm c}$ would be grossly inaccurate; but as these elements
are nearly-free-electron metals in the liquid phase, we expect the
calculation of $\Theta_0^l$ from entropy data to be quite reliable.

\section{Conclusions}
\label{concl}

The theory used here gives us two ways to calculate $\Theta_0^l$,
one using liquid entropy data and valid for all monatomic liquids, and
the other using crystal neutron scattering data and applicable only to
``normal melters,'' for which the crystal and liquid values of
$\Theta_0$ as a function of density are expected to agree
approximately.  This expectation is amply fulfilled by data from 29
normal melters, which show $\Theta_0^{\rm c}$ and $\Theta_0^l$
agreeing within a few percent.  (Our expectation that $\Theta_0^{\rm
c}$ and $\Theta_0^l$ of anomalous melters should strongly disagree is
also verified.)  This work nicely complements calculations by Wallace
and Clements \cite{sod} for pseudopotential sodium, in which they
found at $\rho = 0.925$ g/cm$^3$
\begin{equation}
\Theta_0^{\rm bcc} = 99.65 {\rm \ K \ \ \ and} \ \ \ \Theta_0^l = 98.7 
{\rm \ K},
\label{match}
\end{equation}
which agrees very well with the experimental results for Na from Table
\ref{tabthetas}.  These results, supporting the internal consistency
of this theory, suggest that estimates of $\Theta_0^l$ from liquid
entropy data for anomalous melters, such as those in Table
\ref{tabanom}, will be equally trustworthy; coupled with the
theoretical result given in Eq.\ (\ref{match}), they also show that
theoretical methods have advanced to the point that {\it a priori}
calculations of $\Theta_0^l$, which will be important for the
computation of transport coefficients in the liquid, are also
reliable.
\\ \\ \\
{\bf \Large Appendix: Data and Sources} \\ \\
All of the data used to calculate $\Theta_0^l(\rho_{lm})$ and
$\Theta_0^{\rm c}(\rho_{lm})$ for the 33 elements we consider are
recorded in Table \ref{tabdata}.  The entry in the ``Structure''
column indicates the structure of the crystal at melt; $\rho_{meas}$
is the density at which $\Theta_0^{\rm c}$ was measured; $\gamma^*$ is
the high-$T$ Gr\"{u}neisen parameter; $\rho_{lm}$ is the density of
the liquid at melt at 1 bar (except for Ar; see below); $T_m$ is the
corresponding melting temperature; $S_{expt}^l$ is the experimentally
measured entropy of the liquid at melt; $S_{el}^l$, $S_{anh}^l$, and
$S_{mag}^l$ are the additional contributions to the total entropy from
Eq.\ (\ref{Sl}) ($S_{bdy}^l$ is negligible at $T_m$); and $S_{harm}^l$
is the harmonic contribution to the entropy determined from the other
entropy data using Eq.\ (\ref{Sliq}).  All densities are in g/cm$^3$,
all temperatures are in K, and all entropies are in $k$/atom.  Data in
parentheses are generally less reliable.

\begin{table}
\vspace*{-0.85in}
\hspace*{-1.3in}
\begin{tabular}{llcccccccccc} \hline\hline
Element & Structure & $\rho_{meas}$ & $\Theta_0^{\rm c}(\rho_{meas})$ & 
$\gamma^*$ & $\rho_{lm}$ & $T_m$ & $S_{expt}^l$ & $S_{el}^l$ & $S_{anh}^l$ & 
$S_{mag}^l$ & $S_{harm}^l$ \\ \hline
Cs & bcc   & 1.91   & 27.5  & 1.14  & 1.84   & 301.6 & 11.11   & 0.09   & 0
  & 0    & 11.02 \\
Cd & hcp   & 8.65   & 103   & 2.3   & 8.02   & 594.2 & 9.76    & 0.07   & 0
  & 0    & 9.69  \\
Tl & bcc   & 11.6   & 58.3  & [2]   & 11.22  & 577   & 10.93   & 0.09   & 0
 & 0    & 10.84 \\
Ga & ortho & 5.91   & 162   & 1.5   & 6.09   & 302.9 & 7.18    & 0.04   & 0
  & 0    & 7.14 \\
\\
Si & dia   & 2.34   & 421   & (0.5) & 2.51   & 1687  & 11.05   & 0.22   & 0
  & 0    & 10.83 \\
Ge & dia   & 5.32   & 245   & 0.8   & 5.60   & 1211  & 11.72   & 0.17   & 0
  & 0    & 11.55 \\
Sn & bct   & 7.30   & 103.4 & 2.2   & 7.00   & 505.1 & 9.66    & 0.09   & 0
  & 0    & 9.57 \\
Li & bcc   & 0.546  & 265.5 & 0.88  & 0.515  & 453.7 & 5.66    & 0.04   & 0
 & 0    & 5.62    \\
Na & bcc   & 1.005  & 113.3 & 1.24  & 0.925  & 371.0 & 7.78    & 0.05   & 0
 & 0    & 7.73    \\
K  & bcc   & 0.904  & 68.9  & 1.24  & 0.829  & 336.4 & 9.06    & 0.07   & 0
 & 0    & 8.99    \\
Rb & bcc   & 1.616  & 42.2  & 1.26  & 1.479  & 312.6 & 10.26   & 0.08   & 0
 & 0    & 10.18   \\
Mg & hcp   & 1.74   & 229.4 & 1.5   & 1.59   & 922   & 8.81    & 0.12   & 0
 & 0    & 8.69    \\
Sr & bcc   & 2.6    & 80.0  & [1]   & 2.48   & 1042  & 12.10   & 0.23   & 0
 & 0    & 11.87   \\
Cu & fcc   & 9.018  & 225.3 & 2.02  & 8.00   & 1358  & 10.09   & 0.09   & 0
 & 0    & 10.00   \\
Ag & fcc   & 10.49  & 150.1 & 2.42  & 9.35   & 1235  & 10.94   & 0.10   & 0
 & 0    & 10.84   \\
Au & fcc   & 19.27  & 124.5 & 2.95  & 17.36  & 1338  & 11.77   & 0.11   & 0
 & 0    & 11.66   \\
Zn & hcp   & 7.270  & 161   & 2.2   & 6.58   & 692.7 & 9.04    & 0.07   & 0
 & 0    & 8.97    \\
Hg & rhomb & 14.46  & 64.7  & 2.5   & 13.69  & 234.3 & 8.31    & 0.03   & 0
  & 0    & 8.28    \\
Al & fcc   & 2.731  & 283.5 & 2.25  & 2.385  & 933.5 & 8.58    & 0.11   & 0
 & 0    & 8.47    \\
In & bct   & 7.43   & 85.3  & 2.4   & 7.02   & 429.8 & 9.11    & 0.07   & 0
 & 0    & 9.04    \\
Pb & fcc   & 11.55  & 64.1  & 2.65  & 10.63  & 600.6 & 11.13   & 0.11   & 0
 & 0    & 11.02   \\
\\
Ar & fcc   & 1.7705 & 59.1  & 2.6   & 1.7705 & 219   & 7.33    & 0      & 0
 & 0    & 7.33    \\ 
\\
V  & bcc   & 6.12   & 250   & 1.5   & 5.49   & 2202  & (12.12) & 0.98   & 0
 & 0    & (11.14) \\
Nb & bcc   & 8.58   & 198   & 1.6   & 7.76   & 2744  & 13.24   & 0.94   & 0
 & 0    & 12.30   \\
Ta & bcc   & 16.68  & 162   & 1.6   & 14.8   & 3293  & 14.75   & 1.03   & 0
 & 0    & 13.72   \\
Pt & fcc   & 21.56  & 163.5 & 2.6   & 19.1   & 2045  & 13.09   & 0.85   & 0
 & 0    & 12.24   \\
Pd & fcc   & 12.07  & 200   & 2.3   & 10.5   & 1827  & (12.20) & 0.82   & 0
 & 0    & (11.38) \\
Ni & fcc   & 8.91   & 275   & 1.9   & 7.89   & 1728  & 11.54   & 1.09   & 0
 & 0.31 & 10.14   \\
Cr & bcc   & 7.19   & 338   & (1.5) & (6.28) & 2133  & 11.69   & (0.84) & 
(0.65) & 0.34 & (9.86) \\
Mo & bcc   & 10.22  & 273   & 1.6   & 9.3    & 2896  & (13.26) & 0.67   & 
(0.79) & 0    & (11.80) \\
W  & bcc   & 19.26  & 225   & 1.6   & 17.2   & 3695  & (14.52) & (0.75) & 
(0.65) & 0    & (13.12) \\
\\
Ti & bcc   & 4.24   & 202   & [1]   & 4.14   & 1945  & 11.83   & 1.26   & NA
 & 0    & 10.57   \\
Zr & bcc   & 6.25   & 147   & [1]   & (6.0)  & 2128  & 13.39   & 1.14   & NA
 & 0    & 12.25   \\ 
\hline\hline
\end{tabular}
\caption{The data used to calculate $\Theta_0^l(\rho_{lm})$ and
$\Theta_0^{\rm c}(\rho_{lm})$ for the 33 elements in Tables
\ref{tabthetas} and \ref{tabanom}.  The units of the data, sources and
meanings of the various headings and symbols are discussed in the text
of the Appendix.}
\label{tabdata}
\end{table}

Most of these data are taken from information collected in various
tables in \cite{book}.  For example, the structure information is
taken from Table 19.1, pp.\ 191ff.  Values of $\Theta_0^{\rm
c}(\rho_{meas})$ for materials in the first group were determined
using crystal entropy data collected in Problem 19.8, p.\ 207 (neutron
scattering data is either unavailable or was taken in a phase other
than the one from which the crystal melts); the second, third, and
fourth groups were determined using neutron scattering data recorded
in Table 15.1, p.\ 153; and the final two elements were handled by
taking high-$T$ neutron scattering data from \cite{pet,hei} and
inferring $\Theta_0^{\rm c}$ at $T_m$ (see Figure 9 of \cite{erik}).
These renormalized $\Theta_0^{\rm c}$ include anharmonic
contributions, so $S_{anh}^l$ was neglected for these elements.  (The
fact that $\Theta_0^{\rm c}$ and $\Theta_0^l$ are so close for these
two materials [see Table \ref{tabthetas}] suggests that the crystal
and liquid phases have approximately the same anharmonicity; this is
also suggested by the specific heat data from Figure 1 of \cite{liq}.)
The corresponding values of $\rho_{meas}$ for all materials were
calculated from the experimental temperatures using extensive tables
of density-temperature data at 1 bar for the elements.

The $\gamma^*$ data are also found in Table 19.1 of \cite{book}
(except for the few values in square brackets, which are purely
empirical estimates where data are unavailable), and the $\rho_{lm}$,
$T_m$, and $S_{expt}^l$ data are from Table 21.1, pp.\ 220ff (again,
except for Ar; see the next paragraph).  The electronic entropy
$S_{el}^l$ was determined for the elements in the first two groups
using a free-electron model; $S_{el}^l$ for Ar is zero because Ar is
not a metal; and results for the last two groups were computed using
density of states calculations by Eriksson et al.\ \cite{erik}.
$S_{anh}^l$ was computed, as mentioned in the main text, by assuming
it is equal to $S_{anh}^{\rm c}$, which is negligible for all but the
last five materials in the list (see Tables 19.2 and 19.3, pp.\ 201
and 203 of \cite{book}).  Anharmonicity in the last two is taken into
account by the renormalized $\Theta_0^{\rm c}$ (see the previous
paragraph), and for Cr, Mo, and W, $S_{anh}^{\rm c}$ (and thus
$S_{anh}^l$) was estimated by combining anharmonic entropy
calculations from \cite{erik} and Table 19.3 of \cite{book} to
extrapolate $S_{anh}^{\rm c}$ to the densities in question.  Finally,
$S_{mag}$ for ferromagnetic Ni and antiferromagnetic Cr were also
taken from Table 19.3.

Special treatment was required for Ar because if $\rho_{meas}$ is the
density at which the neutron scattering data were taken and
$\rho_{lm}$ is the density of the liquid at melt at 1 bar,
$|\rho_{meas} - \rho_{lm}|$ is far too large for Eq.\ (\ref{theta0c})
to be reliable.  Thus we decided to set $ \rho_{lm} = \rho_{meas}$, as
shown in Table \ref{tabdata}, and used very extensive tabulated data
of the Ar melt curve to find the pressure such that Ar would melt at
the required density; from this we were able to infer the tabulated
values for $T_m$ and $S_{expt}^l$, and from that point the calculations
proceeded as with the other materials.

Finally, a few words are in order concerning the quality and accuracy
of our data.  On the crystal side, by comparing phonon moments
computed from Born-von K\'{a}rm\'{a}n fits to various different
sources of neutron scattering data, one sees that determinations of
$\Theta_0^{\rm c}(\rho_{meas})$ are typically accurate to 0.5\%; given
that measurements of $\gamma^*$ typically have 10\% error or less, an
additional error of 2\% or so is introduced by the computation of
$\Theta_0^{\rm c}(\rho_{lm})$.  These errors are independent, so the
final error in $\Theta_0^{\rm c}(\rho_{lm})$ lies between 2 and 2.5\%.
For the liquid, comparison of liquid entropy data from separate
high-quality compendia, such as \cite{ch,hult} among others, indicates
errors of $\pm 0.5\%$ for most values of $S_{expt}^l$.  Various
different electronic structure calculations of $S_{el}^l$ vary by
5-10\% at most; these errors will affect our entropy results on about
the 0.1\% level for the materials in the first two groups and 1\% for
those in the last two groups.  Finally, we estimate the error in
$S_{anh}^l + S_{mag}^l$ to be about 0.5\% of $S_{expt}^l$, so the
harmonic part of the liquid entropy $S_{harm}^l$ (and thus
$\Theta_0^l$) will be trustworthy to between 1\% and 1.5\%.  Thus the
error in the ratio is at most 3 to 4\%; this is the source of the
error estimate given in Section \ref{exp}.

Eight materials which have been studied in connection with melting
previously but are not included in this analysis are Kr and Xe, for
which there are no entropy data along the melt curve at sufficiently
high compression (so they cannot be treated as Ar was); and Fe, Th,
Sb, Bi, Ba, and Ca, for which there are no reliable calculations or
estimates of $S_{el}^l$.  This is a significant problem with Fe and
Th, for which $S_{el}^l$ is expected to be large; it is not such a
difficulty with the remaining four, but when combined with various other
problems with their data (for example, neutron scattering was done on
Ca in the fcc phase while the crystal actually melts from bcc), it is
proper to neglect them here as well.

\begin{center} \bf \Large Acknowledgment \end{center}

This work was supported by the U.\ S.\ Department of Energy through 
contract W-7405-ENG-36.


\begin{thebibliography}{99}
\bibitem{diff} E.\ D.\ Chisolm, B.\ E.\ Clements, and D.\ C.\ Wallace, Phys.\
               Rev.\ E {\bf 63}, 031204 (2001); {\bf 64}, 019902 (2001).
\bibitem{march} N.\ H.\ March and M.\ P.\ Tosi, {\it Introduction to Liquid 
                State Physics} (World Scientific, Singapore, 2002), pp.\ 
                134-135.
\bibitem{liq} E.\ D.\ Chisolm and D.\ C.\ Wallace, J.\ Phys.: Condens.\ 
              Matter {\bf 13}, R739 (2001).
\bibitem{wall1} D.\ C.\ Wallace, Phys.\ Rev.\ E {\bf 56}, 4179 (1997).
\bibitem{book} D.\ C.\ Wallace, {\it Statistical Physics of Crystals and
               Liquids} (World Scientific, Singapore, 2003).
\bibitem{sod} D.\ C.\ Wallace and B.\ E.\ Clements, Phys.\ Rev.\ E {\bf 59},
              2942 (1999).
\bibitem{erik} O.\ Eriksson, J.\ M.\ Wills, and D.\ C.\ Wallace, Phys.\ 
               Rev.\ B {\bf 46}, 5221 (1992).
\bibitem{pet} W.\ Petry, A.\ Heiming, J.\ Trampenau, M.\ Alba, C.\ Herzig, 
              H.\ R.\ Schober, and G.\ Vogl, Phys.\ Rev.\ B {\bf 43}, 10933 
              (1991).
\bibitem{hei} A.\ Heiming, W.\ Petry, J.\ Trampenau, M.\ Alba, C.\ Herzig, 
              H.\ R.\ Schober, and G.\ Vogl, Phys.\ Rev.\ B {\bf 43}, 10948 
              (1991).
\bibitem{ch} M.\ W.\ Chase, Jr., C.\ A.\ Davies, J.\ R.\ Downey, Jr., D.\ J.\ 
             Frurip, R.\ A.\ McDonald, and A.\ N.\ Syverud, J.\ Phys.\ Chem.\ 
             Ref.\ Data {\bf 14}, Suppl.\ 1 (1985).
\bibitem{hult} R.\ Hultgren, P.\ D.\ Desai, D.\ T.\ Hawkins, M.\ Gleiser, K.\ 
               K.\ Kelley, and D.\ D.\ Wagman, {\it Selected Values of the 
               Thermodynamic Properties of the Elements} (American Society
               for Metals, Metals Park, Ohio, 1973).
\end{thebibliography}
\end{document}